\catcode`\@=11					



\font\fiverm=cmr5				
\font\fivemi=cmmi5				
\font\fivesy=cmsy5				
\font\fivebf=cmbx5				

\skewchar\fivemi='177
\skewchar\fivesy='60


\font\sixrm=cmr6				
\font\sixi=cmmi6				
\font\sixsy=cmsy6				
\font\sixbf=cmbx6				

\skewchar\sixi='177
\skewchar\sixsy='60


\font\sevenrm=cmr7				
\font\seveni=cmmi7				
\font\sevensy=cmsy7				
\font\sevenit=cmti7				
\font\sevenbf=cmbx7				

\skewchar\seveni='177
\skewchar\sevensy='60


\font\eightrm=cmr8				
\font\eighti=cmmi8				
\font\eightsy=cmsy8				
\font\eightit=cmti8				
\font\eightbf=cmbx8				

\skewchar\eighti='177
\skewchar\eightsy='60


\font\ninei=cmmi9
\font\ninesy=cmsy9

\skewchar\ninei='177
\skewchar\ninesy='60


\font\tenrm=cmr10				
\font\teni=cmmi10				
\font\tensy=cmsy10				
\font\tenex=cmex10				
\font\tenit=cmti10				
\font\tensl=cmsl10				
\font\tenbf=cmbx10				
\font\tentt=cmtt10				
\font\tenss=cmss10				
\font\tensc=cmcsc10				
\font\tenbi=cmmib10				

\skewchar\teni='177
\skewchar\tenbi='177
\skewchar\tensy='60

\def\tenpoint{\ifmmode\err@badsizechange\else
	\textfont0=\tenrm \scriptfont0=\sevenrm \scriptscriptfont0=\fiverm
	\textfont1=\teni  \scriptfont1=\seveni  \scriptscriptfont1=\fivemi
	\textfont2=\tensy \scriptfont2=\sevensy \scriptscriptfont2=\fivesy
	\textfont3=\tenex \scriptfont3=\tenex   \scriptscriptfont3=\tenex
	\textfont4=\tenit \scriptfont4=\sevenit \scriptscriptfont4=\sevenit
	\textfont5=\tensl
	\textfont6=\tenbf \scriptfont6=\sevenbf \scriptscriptfont6=\fivebf
	\textfont7=\tentt
	\textfont8=\tenbi \scriptfont8=\seveni  \scriptscriptfont8=\fivemi
	\def\rm{\tenrm\fam=0 }%
	\def\it{\tenit\fam=4 }%
	\def\sl{\tensl\fam=5 }%
	\def\bf{\tenbf\fam=6 }%
	\def\tt{\tentt\fam=7 }%
	\def\ss{\tenss}%
	\def\sc{\tensc}%
	\def\bmit{\fam=8 }%
	\rm\setparameters\setbaselines\fi}


\font\twelverm=cmr12				
\font\twelvei=cmmi12				
\font\twelvesy=cmsy10	scaled\magstep1		
\font\twelveex=cmex10	scaled\magstep1		
\font\twelveit=cmti12				
\font\twelvesl=cmsl12				
\font\twelvebf=cmbx12				
\font\twelvett=cmtt12				
\font\twelvess=cmss12				
\font\twelvesc=cmcsc10	scaled\magstep1		
\font\twelvebi=cmmib10	scaled\magstep1		

\skewchar\twelvei='177
\skewchar\twelvebi='177
\skewchar\twelvesy='60

\def\twelvepoint{\ifmmode\err@badsizechange\else
	\textfont0=\twelverm \scriptfont0=\eightrm \scriptscriptfont0=\sixrm
	\textfont1=\twelvei  \scriptfont1=\eighti  \scriptscriptfont1=\sixi
	\textfont2=\twelvesy \scriptfont2=\eightsy \scriptscriptfont2=\sixsy
	\textfont3=\twelveex \scriptfont3=\tenex   \scriptscriptfont3=\tenex
	\textfont4=\twelveit \scriptfont4=\eightit \scriptscriptfont4=\sevenit
	\textfont5=\twelvesl
	\textfont6=\twelvebf \scriptfont6=\eightbf \scriptscriptfont6=\sixbf
	\textfont7=\twelvett
	\textfont8=\twelvebi \scriptfont8=\eighti  \scriptscriptfont8=\sixi
	\def\rm{\twelverm\fam=0 }%
	\def\it{\twelveit\fam=4 }%
	\def\sl{\twelvesl\fam=5 }%
	\def\bf{\twelvebf\fam=6 }%
	\def\tt{\twelvett\fam=7 }%
	\def\ss{\twelvess}%
	\def\sc{\twelvesc}%
	\def\bmit{\fam=8 }%
	\rm\setparameters\setbaselines\fi}


\font\fourteenrm=cmr10	scaled\magstep2		
\font\fourteeni=cmmi10	scaled\magstep2		
\font\fourteensy=cmsy10	scaled\magstep2		
\font\fourteenex=cmex10	scaled\magstep2		
\font\fourteenit=cmti10	scaled\magstep2		
\font\fourteensl=cmsl10	scaled\magstep2		
\font\fourteenbf=cmbx10	scaled\magstep2		
\font\fourteentt=cmtt10	scaled\magstep2		
\font\fourteenss=cmss10	scaled\magstep2		
\font\fourteensc=cmcsc10 scaled\magstep2	
\font\fourteenbi=cmmib10 scaled\magstep2	

\skewchar\fourteeni='177
\skewchar\fourteenbi='177
\skewchar\fourteensy='60

\def\fourteenpoint{\ifmmode\err@badsizechange\else
	\textfont0=\fourteenrm \scriptfont0=\tenrm \scriptscriptfont0=\sevenrm
	\textfont1=\fourteeni  \scriptfont1=\teni  \scriptscriptfont1=\seveni
	\textfont2=\fourteensy \scriptfont2=\tensy \scriptscriptfont2=\sevensy
	\textfont3=\fourteenex \scriptfont3=\tenex \scriptscriptfont3=\tenex
	\textfont4=\fourteenit \scriptfont4=\tenit \scriptscriptfont4=\sevenit
	\textfont5=\fourteensl
	\textfont6=\fourteenbf \scriptfont6=\tenbf \scriptscriptfont6=\sevenbf
	\textfont7=\fourteentt
	\textfont8=\fourteenbi \scriptfont8=\tenbi \scriptscriptfont8=\seveni
	\def\rm{\fourteenrm\fam=0 }%
	\def\it{\fourteenit\fam=4 }%
	\def\sl{\fourteensl\fam=5 }%
	\def\bf{\fourteenbf\fam=6 }%
	\def\tt{\fourteentt\fam=7}%
	\def\ss{\fourteenss}%
	\def\sc{\fourteensc}%
	\def\bmit{\fam=8 }%
	\rm\setparameters\setbaselines\fi}


\font\seventeenrm=cmr10 scaled\magstep3		


\newdimen\rp@
\newcount\@basestretchnum
\newskip\@baseskip
\newskip\headskip
\newskip\footskip


\def\setparameters{\rp@=.1em
	\headskip=24\rp@
	\footskip=\headskip
	\delimitershortfall=5\rp@
	\nulldelimiterspace=1.2\rp@
	\scriptspace=0.5\rp@
	\abovedisplayskip=10\rp@ plus3\rp@ minus5\rp@
	\belowdisplayskip=10\rp@ plus3\rp@ minus5\rp@
	\abovedisplayshortskip=5\rp@ plus2\rp@ minus4\rp@
	\belowdisplayshortskip=10\rp@ plus3\rp@ minus5\rp@
	\normallineskip=\rp@
	\lineskip=\normallineskip
	\normallineskiplimit=0pt
	\lineskiplimit=\normallineskiplimit
	\jot=3\rp@
	\setbox0=\hbox{\the\textfont3 B}\p@renwd=\wd0
	\skip\footins=12\rp@ plus3\rp@ minus3\rp@
	\skip\topins=0pt plus0pt minus0pt}


\def\setbaselines{\maxdepth=4\rp@\baselinestretch=\@basestretchnum}


\def\baselinestretch{\afterassignment\@basestretch\@basestretchnum}
\def\@basestretch{%
	\@baseskip=12\rp@ \divide\@baseskip by1000
	\normalbaselineskip=\@basestretchnum\@baseskip
	\baselineskip=\normalbaselineskip
	\bigskipamount=\the\baselineskip
		plus.25\baselineskip minus.25\baselineskip
	\medskipamount=.5\baselineskip
		plus.125\baselineskip minus.125\baselineskip
	\smallskipamount=.25\baselineskip
		plus.0625\baselineskip minus.0625\baselineskip
	\setbox\strutbox=\hbox{\vrule height.708\baselineskip
		depth.292\baselineskip width0pt }}



\def\makeheadline{\vbox to0pt{\baselinestretch=1000
	\vskip-\headskip \vskip1.5pt
	\line{\vbox to\ht\strutbox{}\the\headline}\vss}\nointerlineskip}

\def\makefootline{\baselineskip=\footskip\line{\the\footline}}

\def\big#1{{\hbox{$\left#1\vbox to8.5\rp@ {}\right.\n@space$}}}
\def\Big#1{{\hbox{$\left#1\vbox to11.5\rp@ {}\right.\n@space$}}}
\def\bigg#1{{\hbox{$\left#1\vbox to14.5\rp@ {}\right.\n@space$}}}
\def\Bigg#1{{\hbox{$\left#1\vbox to17.5\rp@ {}\right.\n@space$}}}


\mathchardef\alpha="710B
\mathchardef\beta="710C
\mathchardef\gamma="710D
\mathchardef\delta="710E
\mathchardef\epsilon="710F
\mathchardef\zeta="7110
\mathchardef\eta="7111
\mathchardef\theta="7112
\mathchardef\iota="7113
\mathchardef\kappa="7114
\mathchardef\lambda="7115
\mathchardef\mu="7116
\mathchardef\nu="7117
\mathchardef\xi="7118
\mathchardef\pi="7119
\mathchardef\rho="711A
\mathchardef\sigma="711B
\mathchardef\tau="711C
\mathchardef\upsilon="711D
\mathchardef\phi="711E
\mathchardef\chi="711F
\mathchardef\psi="7120
\mathchardef\omega="7121
\mathchardef\varepsilon="7122
\mathchardef\vartheta="7123
\mathchardef\varpi="7124
\mathchardef\varrho="7125
\mathchardef\varsigma="7126
\mathchardef\varphi="7127
\mathchardef\imath="717B
\mathchardef\jmath="717C
\mathchardef\ell="7160
\mathchardef\wp="717D
\mathchardef\partial="7140
\mathchardef\flat="715B
\mathchardef\natural="715C
\mathchardef\sharp="715D


\def\err@badsizechange{%
	\immediate\write16{--> Size change not allowed in math mode, ignored}}

\baselinestretch=1000
\tenpoint

\catcode`\@=12					

\catcode`\@=11
\expandafter\ifx\csname @iasmacros\endcsname\relax
	\global\let\@iasmacros=\par
\else	\endinput
\fi
\catcode`\@=12


\def\rmb{\seventeenrm}


\def\singlespace{\baselineskip=\normalbaselineskip}
\def\halfspace{\baselineskip=1.5\normalbaselineskip}
\def\doublespace{\baselineskip=2\normalbaselineskip}


\def\AB{\bigskip\parindent=40pt
        \centerline{\bf ABSTRACT}\medskip\halfspace\narrower}
\def\AE{\bigskip\nonarrower\doublespace}
\def\nonarrower{\advance\leftskip by-\parindent
	\advance\rightskip by-\parindent}


\def\boxit#1{\vbox{\hrule\hbox{\vrule\kern3pt
	\vbox{\kern3pt#1\kern3pt}\kern3pt\vrule}\hrule}}

\def\hence{\leavevmode\hbox{\bf .\raise5.5pt\hbox{.}.} }

\def\dalemb#1#2{{\vbox{\hrule height.#2pt
	\hbox{\vrule width.#2pt height#1pt \kern#1pt \vrule width.#2pt}
	\hrule height.#2pt}}}
\def\gtorder{\mathrel{\raise.3ex\hbox{$>$}\mkern-14mu
             \lower0.6ex\hbox{$\sim$}}}
\def\ltorder{\mathrel{\raise.3ex\hbox{$<$}\mkern-14mu
             \lower0.6ex\hbox{$\sim$}}}

\newdimen\fullhsize
\newbox\leftcolumn
\def\twoup{\hoffset=-.5in \voffset=-.25in
  \hsize=4.75in \fullhsize=10in \vsize=6.9in
  \def\fullline{\hbox to\fullhsize}
  \let\lr=L
  \output={\if L\lr
        \global\setbox\leftcolumn=\columnbox\global\let\lr=R \advancepageno
      \else \doubleformat \global\let\lr=L\fi
    \ifnum\outputpenalty>-20000 \else\dosupereject\fi}
  \def\doubleformat{\shipout\vbox{
    \fullline{\box\leftcolumn\hfil\columnbox}\advancepageno}}
  \def\columnbox{\leftline{\vbox{\makeheadline\pagebody\makefootline}}}
  \tolerance=1000 }

\twelvepoint
\doublespace
{\nopagenumbers{
\rightline{~~~November, 1996}
\bigskip\bigskip
\centerline{\rmb A Strategy for a Vanishing Cosmological Constant}
\centerline{\rmb in the Presence of Scale Invariance Breaking}
\medskip
\centerline{\bf Stephen L. Adler
}
\centerline{\bf Institute for Advanced Study}
\centerline{\bf Princeton, NJ 08540}
\medskip}
\bigskip\bigskip
\medskip
\leftline{Second Award in the Gravity Research Foundation Essay Competition 
for 1997; to appear in General Relativity and Gravitation}
\medskip
\leftline{\it Send correspondence to:}
\medskip
{\singlespace\leftline{Stephen L. Adler}
\leftline{Institute for Advanced Study}
\leftline{Olden Lane, Princeton, NJ 08540}
\leftline{Phone 609-734-8051; FAX 609-924-8399; email adler@sns.ias.edu}}
\bigskip\bigskip
}
\vfill\eject
\pageno=2
\AB
Recent work has shown that 
complex quantum field theory emerges as a statistical mechanical  
approximation to 
an underlying noncommutative operator dynamics based on   
a total trace action.  
In this dynamics, scale invariance of the trace action 
becomes the statement $0={\rm Re Tr} T_{\mu}^{\mu}$,  
with $T_{\mu \nu}$ the operator stress energy tensor, and with ${\rm Tr}$ the 
trace over the underlying Hilbert space.  We show that this condition 
implies the vanishing of the cosmological 
constant and vacuum energy in the emergent quantum field theory.  However, 
since the scale invariance condition does not require the operator  
$T_{\mu}^{\mu}$ to vanish, the spontaneous breakdown of 
scale invariance is still permitted.
\AE
\bigskip\bigskip
\vfill\eject
\pageno=3
Perhaps the most baffling problem in current theoretical physics [1] is that 
of understanding the smallness of the observed cosmological 
constant $\Lambda$.  The naive expectation is that one should find 
$\Lambda \sim M_{\rm Planck}^4$, whereas current observational bounds are 120 
orders of magnitude smaller than this, suggesting that there is an exact 
symmetry principle enforcing vanishing of the cosmological constant.  
Unfortunately, in standard quantum field theory no such symmetry principle 
is evident. The two natural candidates are scale invariance and 
supersymmetry, but the empirical facts that particles have rest masses, and 
that bosons and fermions have different mass spectra, tell us that both of 
these symmetries are broken in the observed universe.  It is very difficult  
to understand how either of these symmetries can be broken without the 
breaking communicating itself to the vacuum sector, thereby leading to an 
unacceptably large cosmological constant.  

Let us explicitly illustrate the problem in the case of scale invariance 
symmetry, which is the focus of this essay.  We consider matter fields 
quantized on a background metric $g_{\mu \nu}$, with effective stress energy 
tensor 
operator $T_{eff~\mu \nu}$.  In the limit of a flat background metric, 
Lorentz 
invariance implies that the vacuum expectation of the stress energy tensor 
has the structure 
$$ \langle 0 | T_{eff~\mu \nu} | 0 \rangle = -C g_{\mu \nu} ~~~.\eqno(1)$$
This corresponds to a matter vacuum energy contribution of
$$\langle 0 | T_{eff~0 0} | 0 \rangle = C~~~,\eqno(2)$$
and to a matter-induced contribution to the cosmological constant of  
$$G^{-1}\Lambda_{\rm ind}= 8 \pi  C~~~,\eqno(3)$$
with $G$ Newton's constant.  
Contracting Eq.~(1) with $g^{\mu \nu}$, we can express $C$ in terms of 
the vacuum expectation of the Lorentz index trace of the effective 
matter stress energy tensor, 
giving
$$\eqalign{
C=&-{1 \over 4} \langle 0 |T_{eff~\mu}^{\mu} | 0 \rangle ~~~, \cr
G^{-1}\Lambda_{\rm ind}=&-2 \pi  \langle 0 |T_{eff~\mu}^{\mu} |0 
\rangle~~~.\cr 
}\eqno(4)$$
Suppose now that we lived in an exactly scale invariant world.  Then 
the bare cosmological constant would have to vanish, and since scale 
invariance implies [2] that $T_{eff~\mu}^{\mu}=0$, by Eq.~(4) the matter 
induced 
contribution to the cosmological constant would vanish as well, giving a 
vanishing observed cosmological constant.  But, as we have already noted, 
the assumption of exact scale invariance is not viable:  even if we restrict 
ourselves to theories in which scale invariance anomalies cancel, to be 
relevant to physics these theories must break scale invariance 
so that rest masses are present, leading at least to soft mass terms that 
break the vanishing of $T_{eff~\mu}^{\mu}$.  Part of the problem in trying 
to use scale invariance as a symmetry to enforce vanishing of the 
cosmological constant is that it is a much stronger condition than is 
necessary, since to get the vanishing of a single real number, the 
cosmological 
constant, we must impose the vanishing of an operator $T_{eff~\mu}^{\mu}$, 
a condition equivalent to the vanishing of an infinite number of real numbers.

To seek a way out of this impasse, we turn to a new kinematic framework [3-6]
that we have termed {\it Generalized Quantum Dynamics (GQD)} .  In GQD
the fundamental dynamical variables are symplectic pairs 
of operator-valued variables $\{q_r\},\{p_r\}$ acting on 
an underlying Hilbert space; for simplicity we describe here only 
the case of bosonic operators in a complex Hilbert space, although fermions   
and real and quaternionic Hilbert spaces are readily incorporated into the 
formalism.  The distinguishing feature of GQD is that no {\it a priori} 
commutativity 
properties are assumed for the $q$'s and $p$'s.  Nonetheless, a theory of  
flows in the operator phase space can be set up by focusing on 
{\it total trace functionals}, defined as follows.  Let 
$A[\{q_r\},\{p_r\}]$  be any polynomial in the phase space variables (the 
specification of which depends on giving the ordering of all noncommutative 
factors).  The corresponding real-number valued total trace functional 
${\bf A}[\{q_r\},\{p_r\}]$ is defined as 
$${\bf A}[\{q_r\},\{p_r\}]= {\rm Re Tr}A[\{q_r\},\{p_r\}] 
\equiv {\bf Tr}A[\{q_r\},\{p_r\}]~~~,\eqno(5)$$
where {\rm Tr} denotes the ordinary operator trace.  
We assume sufficient convergence  
for ${\bf Tr}={\rm Re Tr}$ to obey the {\it cyclic property}
$${\bf Tr} {\cal O}_1 {\cal O}_2 = {\bf Tr} {\cal O}_2 {\cal O}_1~~~.\eqno(6)$$

Although noncommutativity of the phase space variables prevents us from 
simply differentiating the operator $A$ with respect to them, we can use 
the cyclic property of {\bf Tr} to define derivatives of the total trace 
functional {\bf A} by forming $\delta {\bf A}$ and cyclically reordering 
all of the operator variations $\delta q_r, \delta p_r$ to the right.
This gives the fundamental definition 
$$\delta {\bf A}={\bf Tr} \sum_r \left( {\delta {\bf A} \over \delta q_r}
\delta q_r + {\delta {\bf A} \over \delta p_r} \delta p_r \right)~~~,
\eqno(7)$$
in which $\delta {\bf A}/\delta q_r$ and $\delta {\bf A}/\delta p_r$
are themselves operators.  Introducing an operator Hamiltonian 
$H[\{q_r\}, \{p_r\}]$ and a corresponding total trace Hamiltonian 
${\bf H}={\bf Tr}H$, the time derivatives of the operator phase space 
variables (denoted by a dot) are
generated by the operator Hamilton equations
$${\delta {\bf H} \over \delta q_r}=-\dot{p_r}~~,~~~   
  {\delta {\bf H} \over \delta p_r}=\dot{q_r}~~.~~~   
\eqno(8)$$
Applying Eq.~(7) to {\bf H} and substituting Eq.~(8), we learn that 
{\bf H} is a constant of the motion.  

Corresponding to the Hamiltonian formalism for GQD there is 
also a Lagrangian formalism following 
from a total trace Lagrangian $\bf{L} [\{q_r\},\{\dot{q_r} \}]$, obtained 
as the Legendre transform of {\bf H}.  Using the Lagrangian formalism, 
repeating the 
standard Noether analysis [3,~4] shows that for a Lagrangian symmetry 
parameterized by a $c$-number parameter $\kappa$, there is a conserved 
{\it total trace} charge ${\bf Q}_{\kappa}$.  Thus, in a Poincar\'e 
invariant theory there is a conserved total trace stress energy tensor 
${\bf T}_{\mu \nu}$, and in a Poincar\'e and scale invariant theory 
the Lorentz index trace of the total trace stress energy tensor vanishes,
$$  {\bf T}_{\mu}^{\mu}=0~~~.\eqno(9)$$
There is, however, no operator analog of Eq.~(9), and this plays a 
crucial role in our argument. 

We have discussed now two apparently unrelated theories: standard 
quantum field theory on the one hand, and the classical, 
noncommutative operator dynamics GQD on the other.  In  recent work with 
Millard [7], we have established  a surprising relation between 
the two, by showing that {\it the statistical mechanics of GQD has a 
structure isomorphic 
to complex quantum field theory}.  The argument is based on the observation 
that in addition to {\bf H}, there are two other generic conserved quantities 
in GQD.  One is the anti-self-adjoint operator
$\tilde C$ defined by  
$$\tilde C=\sum_r [q_r,p_r]~~~,\eqno(10)$$
and the other is the natural integration measure $d \mu$ for the underlying 
operator phase space.  Conservation of $d \mu$  gives a GQD analog 
of Liouville's theorem, and permits the application of statistical 
mechanical methods [7,~8].
For example, 
the canonical ensemble [7] is
given by 
$$\rho=  \rho(\tilde C,\tilde \lambda; {\bf H}, \tau)=    
Z^{-1}\exp(-{\bf Tr} \tilde \lambda \tilde C - \tau {\bf H})~~~,
\eqno(11)$$
with $Z$ a normalization constant (the partition function) 
chosen so that $\int d\mu \rho=1$, 
and with the ensemble parameters $\tau$ and $\tilde \lambda$ (the latter 
an anti-self-adjoint operator) chosen so that the ensemble averages  
$$\langle {\bf H} \rangle_{AV} =\int d\mu \rho {\bf H}~,~~~
\langle \tilde C \rangle_{AV}=\int d\mu \rho \tilde C~~~,\eqno(12)$$
have specified values.  
In general $\langle \tilde C \rangle_{AV}$ can be  
brought to the canonical form
$$\langle \tilde C \rangle_{AV}=i_{eff} D~,~~ i_{eff}=-i_{eff}^{\dagger}~,~~
i_{eff}^2=-1~,~~[i_{eff},D]=0~~~,\eqno(13)$$
with $D$ a real diagonal and non-negative operator. The analysis of Ref.~7
studies the simplest case, in which $D$ is a constant multiple of the unit 
operator in the  underlying Hilbert space, and demonstrates an isomorphism 
between thermodynamic averages in GQD and vacuum expectation values in a
complex quantum field theory, with $i_{eff}$ acting as the imaginary unit 
and with the constant magnitude $D$ playing the role of 
Planck's constant $\hbar$.  
Under this isomorphism, an Hermitian operator ${\cal O}$ 
in GQD corresponds to an effective complex quantum mechanical operator 
$${\cal O}_{eff}={1 \over 2} [{\cal O}-i_{eff} {\cal O} i_{eff}]~~~,\eqno(14)$$
and the ensemble average $\langle {\cal O}_{eff} \rangle_{AV}$ can be 
interpreted as the vacuum expectation $\langle 0|{\cal O}_{eff} |0 \rangle$
in the effective quantum field theory.  

Let us now assume that the observed universe of quantum fields 
is in fact the effective field theory arising from the statistical mechanics 
of an underlying GQD.  Then Eq.~(4) for the matter-induced cosmological 
constant can alternatively be written as a GQD ensemble average
$$G^{-1}\Lambda_{\rm ind}=-2 \pi \langle T_{eff~\mu}^{\mu} \rangle_{AV}
=-2 \pi \int d \mu \rho T_{eff~\mu}^{\mu}~~~.
\eqno(15)$$
Taking the {\bf Tr} of Eq.~(15), and using the relation 
${\bf Tr}{\cal O}_{eff}={\bf Tr} {\cal O}$ which follows from the definition 
of Eq.~(14) and the cyclic property, we get
$$G^{-1}\Lambda_{\rm ind}{\bf Tr}1=-2 \pi \langle {\bf Tr}T_{~\mu}^{\mu} 
\rangle_{AV}~~~.\eqno(16)$$
Since ${\bf Tr}1$ is nonzero (it is the dimension of the underlying Hilbert 
space), comparing Eq.~(16) with Eq.~(9) we see that when the underlying 
GQD is scale invariant, the induced cosmological constant $\Lambda_{\rm ind}$
vanishes.  However, the vanishing of the total trace ${\bf T}_{\mu}^{\mu}
={\rm Re Tr} T_{\mu}^{\mu}$ 
does not require the vanishing of the operator $T_{\mu}^{\mu}$, and so 
the effective quantum field theory can still break scale invariance and 
develop nonzero particle masses.  By the Lorentz invariance argument of   
Eqs.~(1-4), the vanishing of the right hand side of Eq.~(16) also implies 
the vanishing of $\langle {\bf H} \rangle_{AV}$, giving 
a condition [9] relating 
the ensemble parameter $\tau$, which has the dimension of an inverse mass,  
to the dynamically acquired mass scale of the effective theory.

Intuitively, we can describe our picture as follows:  In GQD, chaotic 
motions of the coupled operator degrees of freedom give rise to 
fluctuations, which in the statistical mechanical limit take the form of the  
quantum vacuum fluctuations of the emergent complex quantum field theory.  
Scale invariance in GQD imposes on the underlying fluctuations 
a single constraint,  Eq.~(9),  that at the complex quantum field theory 
level translates into a single restriction on observable parameters, the 
vanishing of the vacuum energy or cosmological constant.  
\bigskip      
\centerline{\bf Acknowledgments}
This work was supported in part by the Department of Energy under
Grant \#DE--FG02--90ER40542.  I wish to thank P. Kumar for comments on the 
initial draft, and to acknowledge the hospitality of the Aspen Center for 
Physics, where this work was begun.  
\vfill\eject
\centerline{\bf References}
\bigskip
\noindent
\item{[1]}  For a good review, see S. Weinberg, Rev. Mod. Phys. {\bf 61}, 1 
(1989).
\bigskip 
\noindent
\item{[2]}  See, e.g., S. Coleman, {\it Aspects of symmetry}, 
Cambridge University Press, Cambridge, 1985, Chapt. 3.
\bigskip
\noindent
\item{[3]}  S. L. Adler, Nucl. Phys. {\bf B 415}, 195 (1994).
\bigskip
\noindent
\item{[4]}  S. L. Adler, {\it Quaternionic quantum mechanics and quantum 
fields},
Oxford, New York, 1995, Secs. 13.5-13.7 and App. A.
\bigskip
\noindent
\item{[5]}  S. L. Adler, G. V. Bhanot, and J. D. Weckel, J. Math. Phys. {\bf 
35}, 
531 (1994).
\bigskip
\noindent
\item{[6]}  S. L. Adler and Y.-S. Wu, Phys. Rev. {\bf D 49}, 6705 (1994).
\bigskip
\noindent
\item{[7]}  S. L. Adler and A. C. Millard, Nuc. Phys. {\bf B 473}, 199 (1996).
\bigskip
\noindent
\item{[8]}  S. L. Adler and L. P. Horwitz, ``Microcanonical ensemble and 
algebra 
of conserved generators for generalized quantum dynamics'', J. Math. Phys. 
(in press).
\bigskip
\noindent
\item{[9]}  When fermions are included as in Ref.~7, scale invariance implies 
both ${\rm Re Tr}T_{\mu}^{\mu}=0$ and ${\rm Re Tr} (-1)^F  T_{\mu}^{\mu}=0$.  
Hence by covariance $\langle {\rm Re Tr}H \rangle_{AV}=
\langle {\rm Re Tr}(-1)^F H \rangle_{AV}=0$, giving conditions on 
the ensemble parameters $\hat \tau$ and $\tau$.

\vfill
\eject
\bye